# Large thickness dependence of the carrier mobility in a transparent oxide semiconductor, La-doped BaSnO$_3$

Anup V. Sanchela[1*], Mian Wei[2], Haruki Zensyo[3], Bin Feng[4], Joonhyuk Lee[5], Gowoon Kim[5], Hyoungjeen Jeen[5], Yuichi Ikuhara[4], and Hiromichi Ohta[1,2*]

[1]*Research Institute for Electronic Science, Hokkaido University, N20W10, Kita, Sapporo 001−0020, Japan*
[2]*Graduate School of Information Science and Technology, Hokkaido University, N14W9, Kita, Sapporo 060−0814, Japan*
[3]*School of Engineering, Hokkaido University, N14W9, Kita, Sapporo 060−0814, Japan*
[4]*Institute of Engineering Innovation, The University of Tokyo, 2−11−16 Yayoi, Bunkyo, Tokyo 113−8656, Japan*
[5]*Department of Physics, Pusan National University, Busan 46241, Korea*

Correspondence and requests for materials should be addressed to:
A.S. (anup.sanchela@es.hokudai.ac.jp) and H.O. (hiromichi.ohta@es.hokudai.ac.jp)

**In this study, we report that the carrier mobility of 2%-La-doped BaSnO$_3$ (LBSO) films on (001) SrTiO$_3$ and (001) MgO substrates strongly depends on the thickness whereas it is unrelated to the film/substrate lattice mismatch (+5.4 % for SrTiO$_3$, −2.3 % for MgO). The films exhibited large differences in the lattice parameters, the lateral grain sizes (~85 nm for SrTiO$_3$, ~20 nm for MgO), the surface morphologies, the threading dislocation densities, and the misfit dislocation densities. However, the mobility dependences on the film thickness in both cases were almost the same, saturating at ~100 cm$^2$ V$^{-1}$ s$^{-1}$ while the charge carrier densities approached the nominal carrier concentration (=[2 % La$^{3+}$]). Our study clearly indicates that the carrier mobility in LBSO films strongly depends on the thickness. These results would be beneficial for understanding the carrier transport properties and fruitful to further enhance the mobility of LBSO films.**

Transparent oxide semiconductors (TOSs) showing high optical transparency and high electrical conductivity have been applied as active materials in wide-scale advanced electronic device applications[1,2]. Recently, there has been growing interest on La-doped BaSnO$_3$ (LBSO, bandgap, $E_g \sim$ 3.5 eV) with cubic perovskite structure ($a$ = 4.115 Å) as a novel TOS because flux-grown LBSO single crystals exhibited a very high mobility of 320 cm$^2$ V$^{-1}$ s$^{-1}$ (carrier concentration 8 × 10$^{19}$ cm$^{-3}$) at room temperature (RT)[3,4]. Such high mobility is originated from its small carrier effective mass ($m^*$ = 0.40 $m_e$[5]) and long carrier relaxation time[6]. Therefore, many researchers have tried to prepare high mobility LBSO epitaxial films to date, but the observed mobility has been low compared to that of single crystals[7-16].

Recently, Paik *et al*. obtained the highest mobility of 183 cm$^2$ V$^{-1}$ s$^{-1}$ in a LBSO film on DyScO$_3$ ($a$ = 3.943 Å, $\Delta a$ = +4.2 %)[7]. Raghavan *et al*. also achieved high mobilities of 150 cm$^2$ V$^{-1}$ s$^{-1}$ and 100 cm$^2$ V$^{-1}$ s$^{-1}$ in LBSO films deposited by high purity molecular beam epitaxy (MBE) technique on PrScO$_3$ ($a$ = 4.026 Å, $\Delta a$ = +2.18 %) and SrTiO$_3$, respectively[8]. Higgins *et al*. reported mobility values up to 81 cm$^2$ V$^{-1}$ s$^{-1}$ in LBSO films grown on TbScO$_3$ (110) ($a$ = 3.958 Å, $\Delta a$ = +3.97 %) by MBE[9]. A low mobility value of 10 cm$^2$ V$^{-1}$ s$^{-1}$ was also reported by Wadekar *et al*. in a LBSO film on SmScO$_3$ (110) ($a$ = 3.991 Å, $\Delta a$ = +3.1 %)[10]. These studies attributed the origin of the low mobility to the misfit/threading dislocations, which are generated from a large lattice mismatch ($\Delta a$) at the film/substrate interface (i.e. $\Delta a$ = +5.4 % for LBSO/SrTiO$_3$ interface)[11-14].

In order to minimize $\Delta a$, a buffer layer deposition on a substrate was also investigated[15-20]. Shin *et al*. used undoped BaSnO$_3$ film as a buffer layer (150 nm thick) on MgO by pulsed laser deposition (PLD) technique and obtained a mobility of 97.2 cm$^2$ V$^{-1}$ s$^{-1}$ [17]. Another study by Shiogai *et al*. reported a mobility of 80 cm$^2$ V$^{-1}$ s$^{-1}$ with (Sr, Ba)SnO$_3$ buffer (200-nm-thick) deposited by PLD on SrTiO$_3$[15]. Lee *et al*. have used the flux grown undoped BaSnO$_3$ (001) single crystal as a substrate, but the resulting mobilities were < 100 cm$^2$ V$^{-1}$ s$^{-1}$[16]. These contradict the hypothesis regarding the misfit/threading dislocations since there were almost no lattice mismatch between the substrate and the film.

Several studies suggest that cation off-stoichiometry or cation mixing can introduce charge point defects[21-23] and dislocations, which act as scattering sources and thus suppress the mobility. However, the origin of the limited electron mobility in LBSO thin films has not been clearly explained to date, and a fundamental study on other factors such as the film thickness is required to understand this phenomenon. Therefore, in this study, we analyzed the structural and electrical features of epitaxial LBSO ($La_{0.02}Ba_{0.98}SnO_3$) films with various thicknesses (14−1040 nm), which were grown on (001) perovskite $SrTiO_3$ ($\Delta a$ = +5.4 %) and non-perovskite (001) MgO ($\Delta a$ = −2.3 %) by PLD.

Here we report that the carrier mobility of the LBSO films strongly depends on the thickness whereas it is unrelated to the lattice mismatch. Although we observed large differences in lattice parameters, lateral grain size, density of threading dislocations, surface morphology, and density of misfit dislocations, the mobility increased almost simultaneously with the thickness in both cases and saturated at ~100 $cm^2$ $V^{−1}$ $s^{−1}$, together with the approaching to the nominal carrier concentration (= [2% $La^{3+}$]), clearly indicating that the behavior of mobility depends on the film thickness. The present results would be beneficial to understand the behavior of mobility and fruitful to further enhance the mobility of La-doped $BaSnO_3$ thin films.

Epitaxial $La_{0.02}Ba_{0.98}SnO_3$ films with thicknesses varying from 14 nm to 1.04 μm were heteroepitaxially grown on (001) $SrTiO_3$ and (001) MgO single crystal substrates by PLD technique using a KrF excimer laser ($\lambda$ = 248 nm, fluence ~ 2 J $cm^{−2}$ $pulse^{−1}$, 10 Hz). The temperatures during the film growth were 700 °C for $SrTiO_3$ substrates and 750 °C for MgO substrates while the oxygen pressure was kept at 10 Pa. In case of $SrTiO_3$ substrates, the LBSO films were annealed at 1200 °C in the air to obtain atomically smooth surfaces.[5, 15]

High-resolution X-ray diffraction (Cu K$\alpha_1$, ATX-G, Rigaku Co.) measurements revealed that the LBSO films were heteroepitaxially grown on (001) $SrTiO_3$ substrates and (001)

MgO substrates with a cube-on-cube epitaxial relationship. The film thicknesses were determined from the Kiessig fringes or Pendelloesung fringes. Surface morphology was investigated by an atomic force microscopy (AFM, Nanocute, Hitachi High Tech.). Stepped and terraced surface was observed on the film grown on (001) SrTiO$_3$ substrate **[Supplementary Fig. S1(a)]** whereas very tiny grains were observed in the film grown on (001) MgO substrate **[Supplementary Fig. S1(b)]**. The AFM images show that the films grown on SrTiO$_3$ and MgO have very different surface morphologies.

In order to elaborate the structural differences in more detail, X-ray reciprocal space mappings (RSMs) were performed around the asymmetric 103 diffraction spot of BaSnO$_3$ with the 103 diffraction spot of SrTiO$_3$ **[Fig. 1(a)]** and the 204 diffraction spot of MgO **[Fig. 1(b)]**, respectively. While the $q_x/2\pi$ peak position of BaSnO$_3$ and the substrate are different from each other, they are both located nearby the red dotted line (cubic), indicating incoherent epitaxial growth occurred in both cases. In order to determine the lateral grain size ($D$), we plotted the cross-sectional peak intensity as a function of $q_x/2\pi$. In case of the SrTiO$_3$ substrate, an integral width of 0.0306 nm$^{-1}$ was obtained for the 14-nm-thick film whereas that of 0.0127 nm$^{-1}$ was obtained for the 1040-nm-thick film **[Fig. 1(c)]**. In case of the MgO substrate, an integral width of 0.1085 nm$^{-1}$ was obtained for the 44-nm-thick film whereas that of 0.0557 nm$^{-1}$ was obtained for the 1000-nm-thick film **[Fig. 1(d)]**.

Using the 103 diffraction spots, we calculated the average lattice parameters, $(a^2 \cdot c)^{1/3}$, of the LBSO films grown on SrTiO$_3$ and MgO substrates **[Fig. 2(a)]**, where *a* and *c* are the in-plane and the out-of-plane lattice parameters, respectively. The $(a^2 \cdot c)^{1/3}$ values of the films on SrTiO$_3$ and MgO substrates initially showed opposite behaviors; $(a^2 \cdot c)^{1/3}$ of the films on MgO was larger than the bulk whereas that of the films on SrTiO$_3$ was smaller than the bulk, which are probably attributed to the differences in the lattice mismatch. On both substrates, the $(a^2 \cdot c)^{1/3}$ values were nearly similar when the thickness was greater than 300 nm. We then calculated the lateral grain size [$D$ = (integral width in $q_x/2\pi$ direction of the RSM)$^{-1}$] of the LBSO films grown on SrTiO$_3$ and MgO substrates as shown in **Fig. 2(b)**. The lateral grain sizes were quite different as the LBSO films on SrTiO$_3$ exhibited a maximum grain size of ~85 nm whereas the grains in the LBSO films on MgO were 20 nm or less.

The microstructure of the LBSO films was characterized by HAADF-STEM. **Figure 3** shows cross-sectional HAADF-STEM images of (a) the 1.04-μm-thick LBSO/SrTiO$_3$ and (b) the 1-μm-thick LBSO/MgO films. In case of LBSO/SrTiO$_3$ **[Fig. 3(a)]**, mismatch dislocations (indicated by arrows) are observed periodically at the interface. The spacing of the mismatch dislocation was about 7.3 nm, which is in good agreement with the Δ*a* = +5.3 %. On the other hand, in case of LBSO/MgO **[Fig. 3(b)]**, mismatch dislocations were not periodic but seemingly occasional. Further, we observed high density threading dislocations in the cross-sectional LAADF-STEM images of the films as shown in **Supplementary Fig. S2**. The average distance between two threading dislocations is ~100 nm for LBSO/SrTiO$_3$ [**Fig. S2(a)**] and ~30 nm for LBSO/MgO [**Fig. S2(b)**], reflecting the lateral grain sizes obtained from the RSMs (~85 nm for LBSO/SrTiO$_3$ and ~20 nm for LBSO/MgO). Thus, the densities of the threading dislocations are $1.4 \times 10^{10}$ cm$^{-2}$ for the film on SrTiO$_3$ substrate and $2.5 \times 10^{11}$ cm$^{-2}$ for the film on MgO substrate. These results show that there are several structural differences between the LBSO films on SrTiO$_3$ and MgO substrates, which include surface morphology, lattice parameter, lateral grain size, density of threading dislocations, and density of misfit dislocations.

The electrical resistivity (*ρ*), carrier concentration (*n*), and Hall mobility ($\mu_{Hall}$) and thermopower (*S*) of the LBSO films at room temperature (RT) and low temperatures were measured by the conventional DC four-probe method using an In-Ga alloy electrode with van der Pauw geometry. *S*-values were measured by creating a temperature difference (Δ*T*) of ~4 K across the film using two Peltier devices. Two small thermocouples were used to monitor the actual temperatures of each end of the film. The thermo-electromotive force (Δ*V*) and Δ*T* were measured simultaneously, and the *S*-values were obtained from the slope of the Δ*V*–Δ*T* plots (correlation coefficient > 0.9999).

**Figure 4** and **Supplementary Table I−II** summarize the electron transport properties of the LBSO films grown on SrTiO$_3$ and MgO substrates at RT. Regarding the overall tendencies, no clear difference was observed from LBSO films deposited on SrTiO$_3$ and MgO substrates. The value of *n* increased with increasing thickness and approached the nominal carrier concentration (= [2% La$^{3+}$]). Approximately, 88 % La$^{3+}$ dopants were activated and produced conducting electrons for films thicker than 350 nm **[Fig. 4(a)]**. Similarly, the magnitude of *S* **[Fig. 4(b)]**, which decreases with increasing *n*, gradually decreased with the thickness, which is consistent with **Fig. 4(a)**. All values of *S* were negative, indicating the LBSO films are *n*-type semiconductors[24]. In addition, $\mu_{Hall}$ increased gradually with thickness and became constant for films thicker than 350 nm. The highest mobility values were 97.7 cm$^2$ V$^{-1}$ s$^{-1}$ for 1040 nm thick LBSO/SrTiO$_3$ and 99.2 cm$^2$ V$^{-1}$ s$^{-1}$ for 450 nm thick LBSO/MgO. The thickness dependence of $\mu_{Hall}$ and *n* in the LBSO films were similar on both SrTiO$_3$ and MgO substrates. In addition, since $\mu_{Hall}$ and *n* of LBSO films thicker than 350 nm do not show a significant dependence on the substrates, the contributions from the structural differences between LBSO/SrTiO$_3$ and LBSO/MgO on the mobility are likely small.

The electron mobility in LBSO films rapidly increased with the thickness. However, the maximum mobility (~100 cm$^2$ V$^{-1}$ s$^{-1}$) was still low compared to the bulk values (~320 cm$^2$ V$^{-1}$ s$^{-1}$). In order to further investigate the suppression of electronic transport properties, we performed x-ray absorption spectroscopy (XAS) around the Sn M$_{4,5}$ edge of a 500-nm-thick LBSO film on SrTiO$_3$ substrate in Pohang accelerator laboratory (2A) (**See Supplementary Fig. S3**). Several peak structures (A–F) were clearly observed in the XAS spectra. The peaks labelled as B–F are well matched with BaSnO$_3$. However, there is an additional peak from 2+ valence state of Sn (SnO, peak A). Since Sn$^{2+}$ ions in LBSO films should play as electron acceptors as well as ionized impurities, they may be related to the suppression of $\mu_{Hall}$ and *n* in the films (compared to bulk values).

Finally, we measured the temperature dependence of the electron transport properties of the LBSO films grown on (001) SrTiO$_3$ substrates (**See Supplementary Fig. S4**).

Metallic behavior was observed in all the films [**Fig. S4(a)**], indicating that the Fermi energy is located above the conduction band edge and the films behave as degenerate semiconductors[14]. The values of $n$ for all films were almost temperature independent and similar for films thicker than 350 nm [**Fig. S4(b)**]. $\mu_{Hall}$ increased with decreasing temperature, and the change was more significant for thicker films. The highest $\mu_{Hall}$ of 163 cm$^2$ V$^{-1}$ s$^{-1}$ was observed in 1040 nm thick LBSO film at 8 K [**Fig. S4(c)**]. $|S|$ almost linearly decreased with decreasing temperature [**Fig. S4(d)**], which is typical for degenerate semiconductors[24].

In summary, we have demonstrated that the electron transport properties of the LBSO films grown on (001) SrTiO$_3$ and (001) MgO substrates show strong thickness dependence in the range of 14 nm to 1040 nm. Although the LBSO/SrTiO$_3$ and LBSO/MgO exhibited several structural differences including lattice parameters, lateral grain size, density of threading dislocations, surface morphology, and density of misfit dislocations, these structural discrepancies did not play a major role in the carrier mobility as no clear structure-induced difference was observed in the electron transport properties. $\mu_{Hall}$ and $n$ increased with increasing LBSO film thickness. On both SrTiO$_3$ and MgO substrates, the maximum $\mu_{Hall}$ observed was ~100 cm$^2$ V$^{-1}$ s$^{-1}$. While the origin of the strong thickness dependence of $\mu_{Hall}$ remains unclear, we detected 2+ valence state of Sn in the XAS spectrum of a 500 nm thick LBSO film. Since Sn$^{2+}$ ions should play not only as electron acceptors but also ionized impurities, they may increase the scattering cross section of the electrons and contribute to the mobility suppression[21-23]. We hope to clarify the effect of Sn valence state on the electron mobility of LBSO films in near future.

We believe that our results can provide a guideline for the thickness optimization of high-mobility LBSO films grown on other substrates. This study also provides further insights on the development of LBSO-based electronic devices.

**Acknowledgments**
This research was supported by Grants-in-Aid for Scientific Research on Innovative

Areas "Nano Informatics" (25106003 and 25106007) from the Japan Society for the Promotion of Science (JSPS). H.J. and H.O. are supported by the Korea-Japan bilateral program funded from following programs of each country: International cooperation program by the NRF (2018K2A9A2A08000079, FY2018) and JSPS. H.O. was supported by Grants-in-Aid for Scientific Research A (17H01314) from the JSPS, the Asahi Glass Foundation, and the Mitsubishi Foundation. A part of this work was supported by Dynamic Alliance for Open Innovation Bridging Human, Environment, and Materials, and by the Network Joint Research Center for Materials and Devices.

**Contributions**

A.S., M.W., and H.Z. performed the sample preparation and electron transport measurements. B.F. and Y.I. performed the STEM analyses. J.L., G.K., and H.J. performed the XAS measurements. H.O. planned and supervised the project. All authors discussed the results and commented on the manuscript.

**Competing financial interests**

The authors declare no competing financial interests.

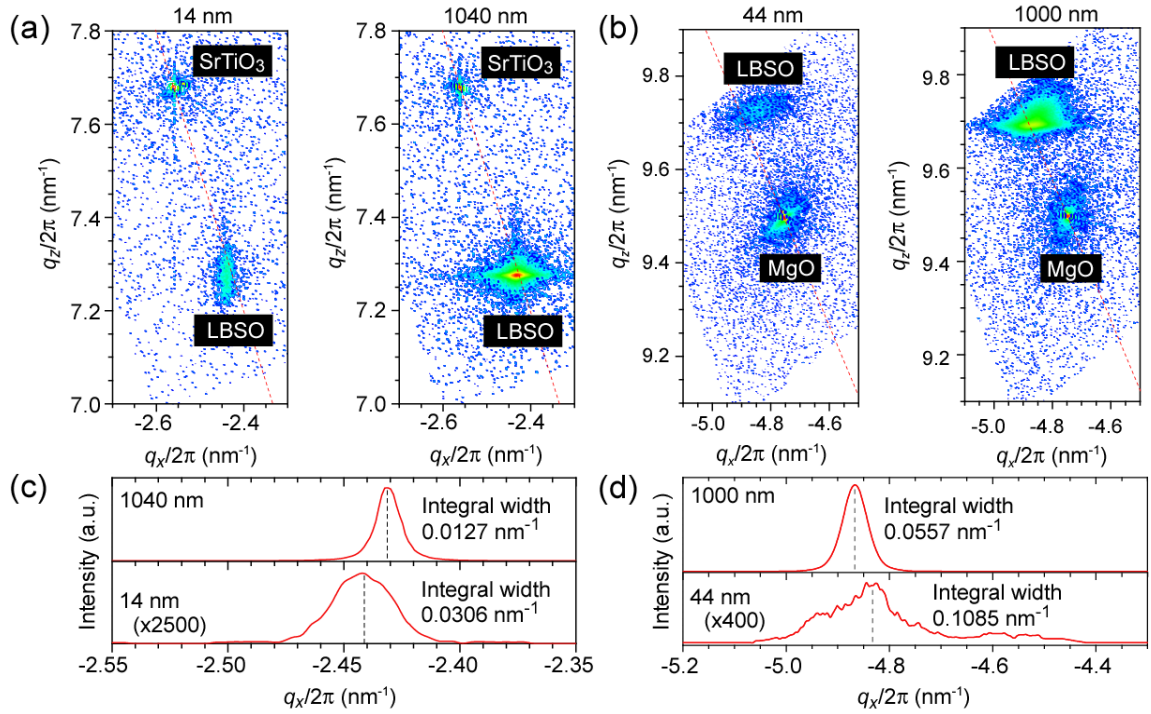

**FIG. 1 | Crystallographic characterization of the LBSO epitaxial films grown on (001) SrTiO₃ and (001) MgO substrates.** X-ray reciprocal space mapping around 103 LBSO on (a) (001) SrTiO$_3$ substrates (14-nm-thick and 1040-nm-thick films) and (b) (001) MgO substrates (44-nm-thick and 1000-nm-thick films). The red dotted lines show cubic symmetry. Cross-sectional intensity profiles of 103 LBSO peak on (c) (001) SrTiO$_3$ substrates (from (a)) and (d) (001) MgO substrates (from (b)).

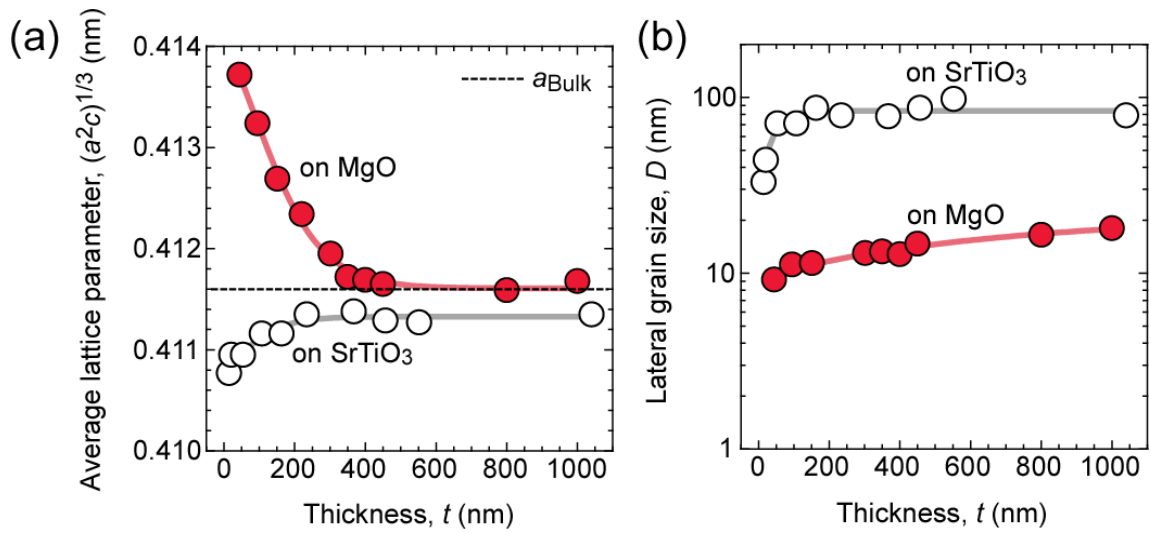

**FIG. 2** | **Thickness dependence of the crystal quality of the LBSO epitaxial films grown on (001) SrTiO$_3$ and (001) MgO substrates.** (a) Average lattice parameters $(a^2c)^{1/3}$ and (b) lateral grain size (*D*) of the LBSO films grown on SrTiO$_3$ (white) and MgO (red). Large differences in the lattice parameters and the lateral grain sizes were observed.

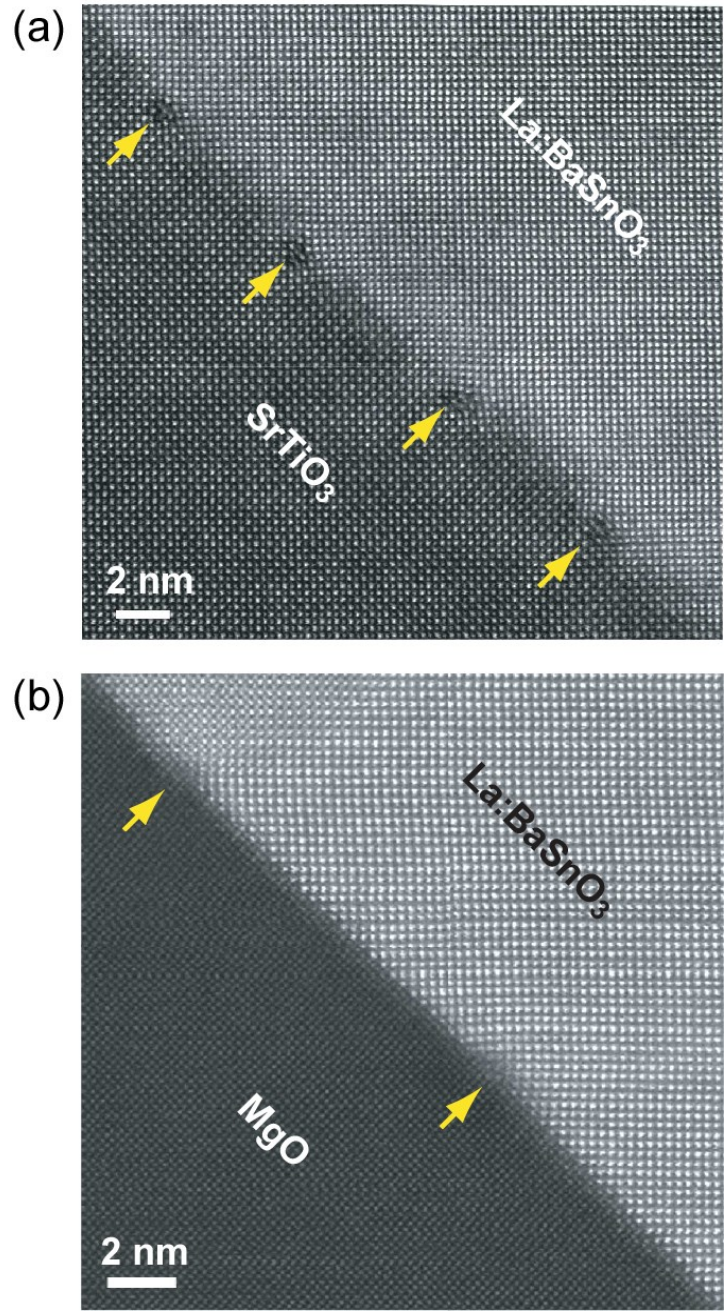

**FIG. 3 | Cross-sectional HAADF-STEM images for the LBSO epitaxial films.** (a) 1040-nm-thick LBSO film grown on SrTiO$_3$ substrate. (b) 1000-nm-thick LBSO film grown on MgO substrate. While mismatch dislocations (arrow) are seen periodically (~7.3 nm) in (a), such periodicity is not clearly seen in (b), indicating the difference of the mismatch dislocation density

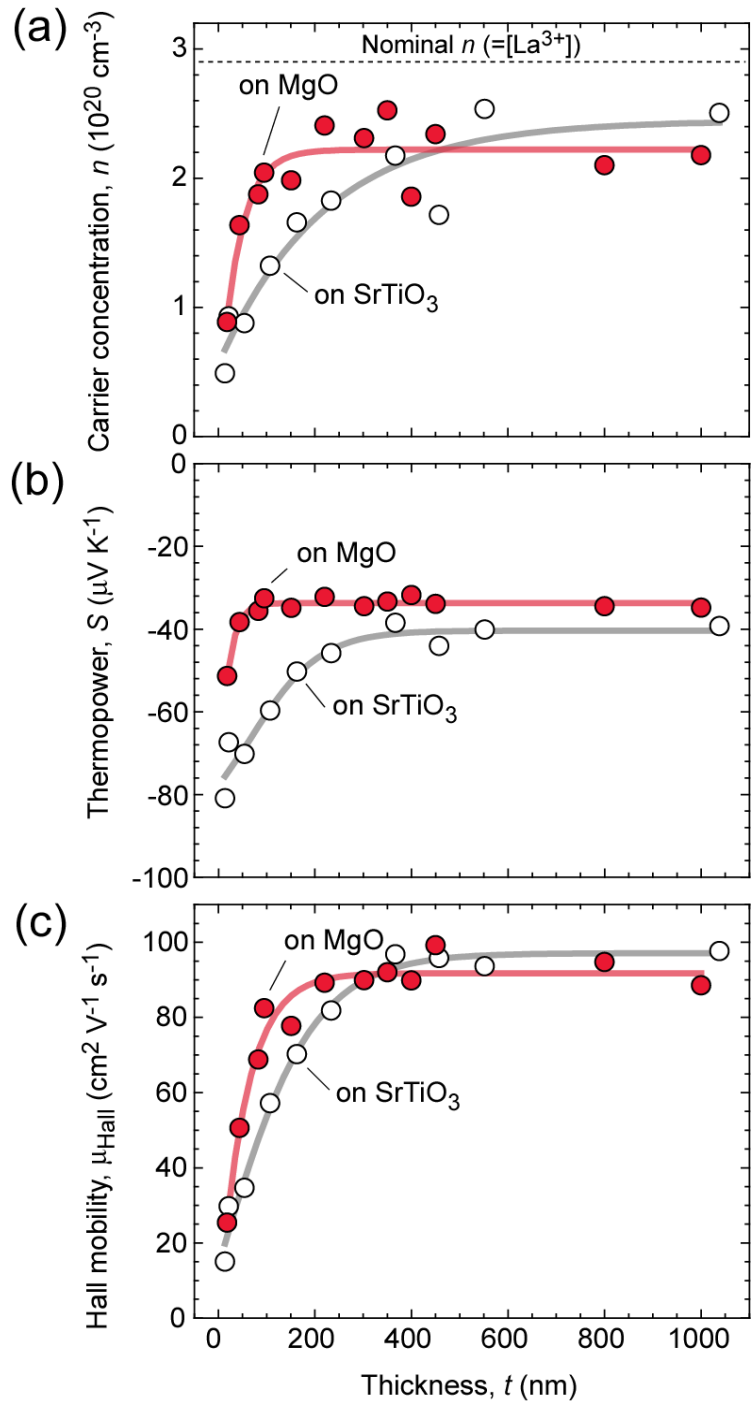

**FIG. 4 | Thickness dependent electron transport properties of the LBSO epitaxial films at RT.** [(a) carrier concentration ($n$), (b) thermopower ($S$), and (c) Hall mobility ($\mu_{Hall}$)]. No clear differences in the overall tendencies were observed on SrTiO$_3$ and MgO substrates.

# Supplementary Information

# Large thickness dependence of the carrier mobility in a transparent oxide semiconductor, La-doped BaSnO$_3$

Anup V. Sanchela[1*], Mian Wei[2], Haruki Zensyo[3], Bin Feng[4], Joonhyuk Lee[5], Gowoon Kim[5], Hyoungjeen Jeen[5], Yuichi Ikuhara[4], and Hiromichi Ohta[1,2*]

[1]*Research Institute for Electronic Science, Hokkaido University, N20W10, Kita, Sapporo 001−0020, Japan*
[2]*Graduate School of Information Science and Technology, Hokkaido University, N14W9, Kita, Sapporo 060−0814, Japan*
[3]*School of Engineering, Hokkaido University, N14W9, Kita, Sapporo 060−0814, Japan*
[4]*Institute of Engineering Innovation, The University of Tokyo, 2−11−16 Yayoi, Bunkyo, Tokyo 113−8656, Japan*
[5]*Department of Physics, Pusan National University, Busan 46241, Korea*

Correspondence and requests for materials should be addressed to:
A.S. (anup.sanchela@es.hokudai.ac.jp) and H.O. (hiromichi.ohta@es.hokudai.ac.jp)

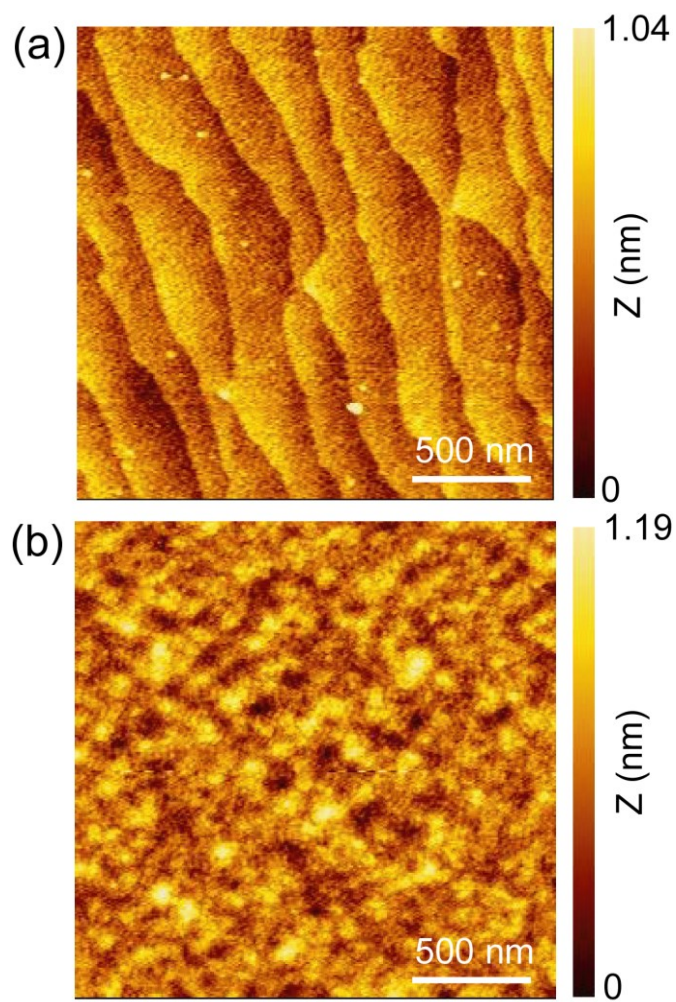

**Supplementary FIG. S1 | Topographic AFM images of the LBSO films grown on (a) (001) SrTiO₃ and (b) (001) MgO substrates.** Stepped and terraced surface is observed in (a) whereas very tiny grains were observed in (b). The topographic AFM images of the LBSO films show different surface morphologies on the two different substrates.

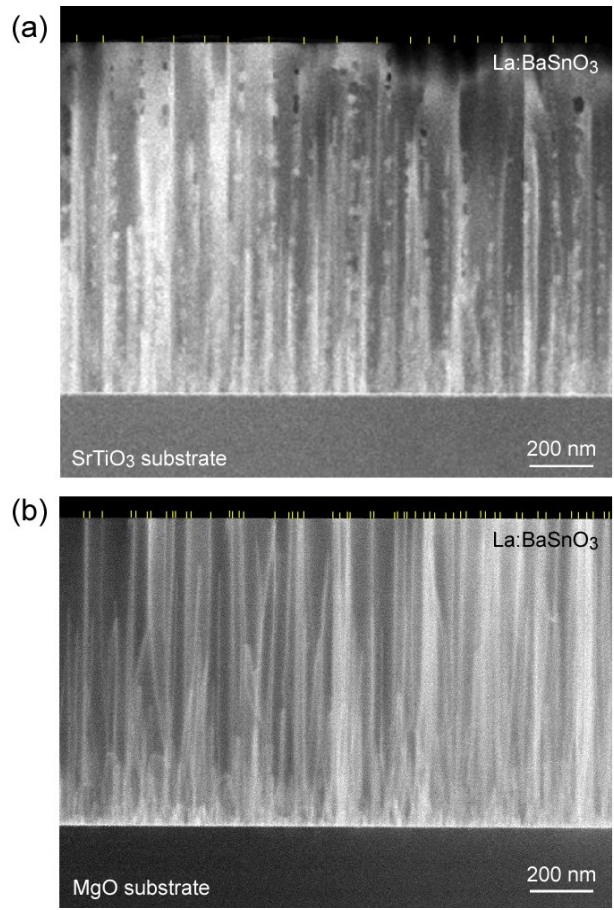

**Supplementary FIG. S2 | Cross-sectional LAADF-STEM images of the LBSO films grown on (a) (001) SrTiO$_3$ and (b) (001) MgO substrates.** Columnar structures are seen in both cases. The threading dislocation densities were roughly estimated using cross-sectional LAADF-STEM, which is sensitive to strain. Threading dislocations at the film surface are shown as yellow bars. The average distances between two threading dislocations were ~100 nm for (a) LBSO/SrTiO$_3$ and ~30 nm for (b) LBSO/MgO, reflecting the lateral grain size obtained from the RSMs (~80 nm for LBSO/SrTiO$_3$ and ~20 nm for LBSO/MgO).

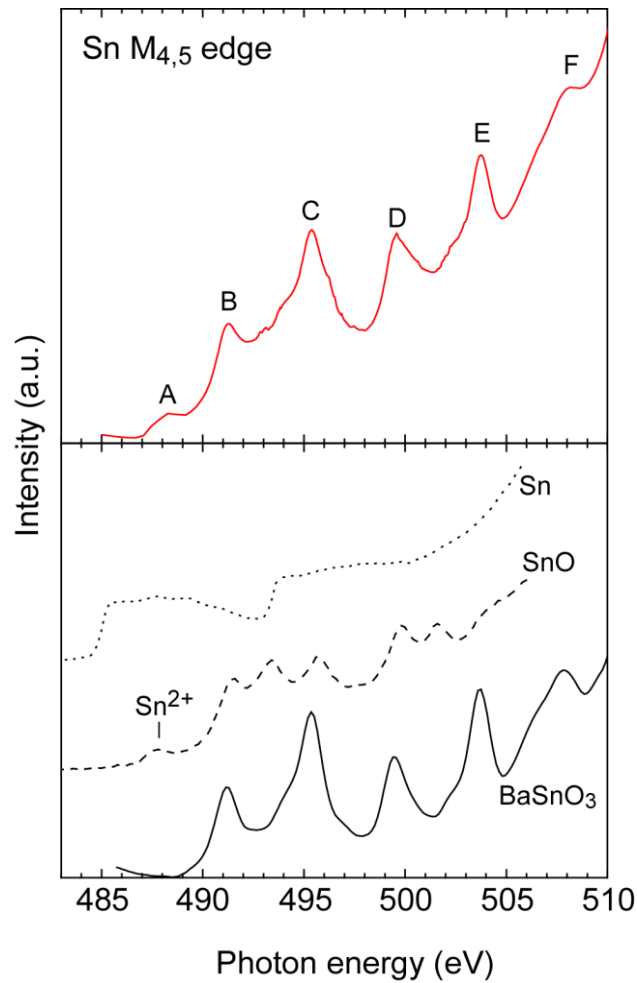

**Supplementary FIG. S3** | (Upper) XAS spectrum around Sn $M_{4,5}$ edge of the 500-nm-thick LBSO film grown on (001) $SrTiO_3$ substrate (This work). (Lower) Reported XANES (X-ray absorption near-edge structure spectroscopy) spectra of Sn foil [1] and SnO powder [1], and HXPES (Hard X-ray photoelectron spectra) spectrum of slightly La-doped $BaSnO_3$ epitaxial film on (110) $TbScO_3$ [2]. Several peak structures (A–F) were clearly observed in the XAS spectra. The peaks of B–F are well matched with $BaSnO_3$. However, we have also detected 2+ valence state of Sn (SnO, peak A).

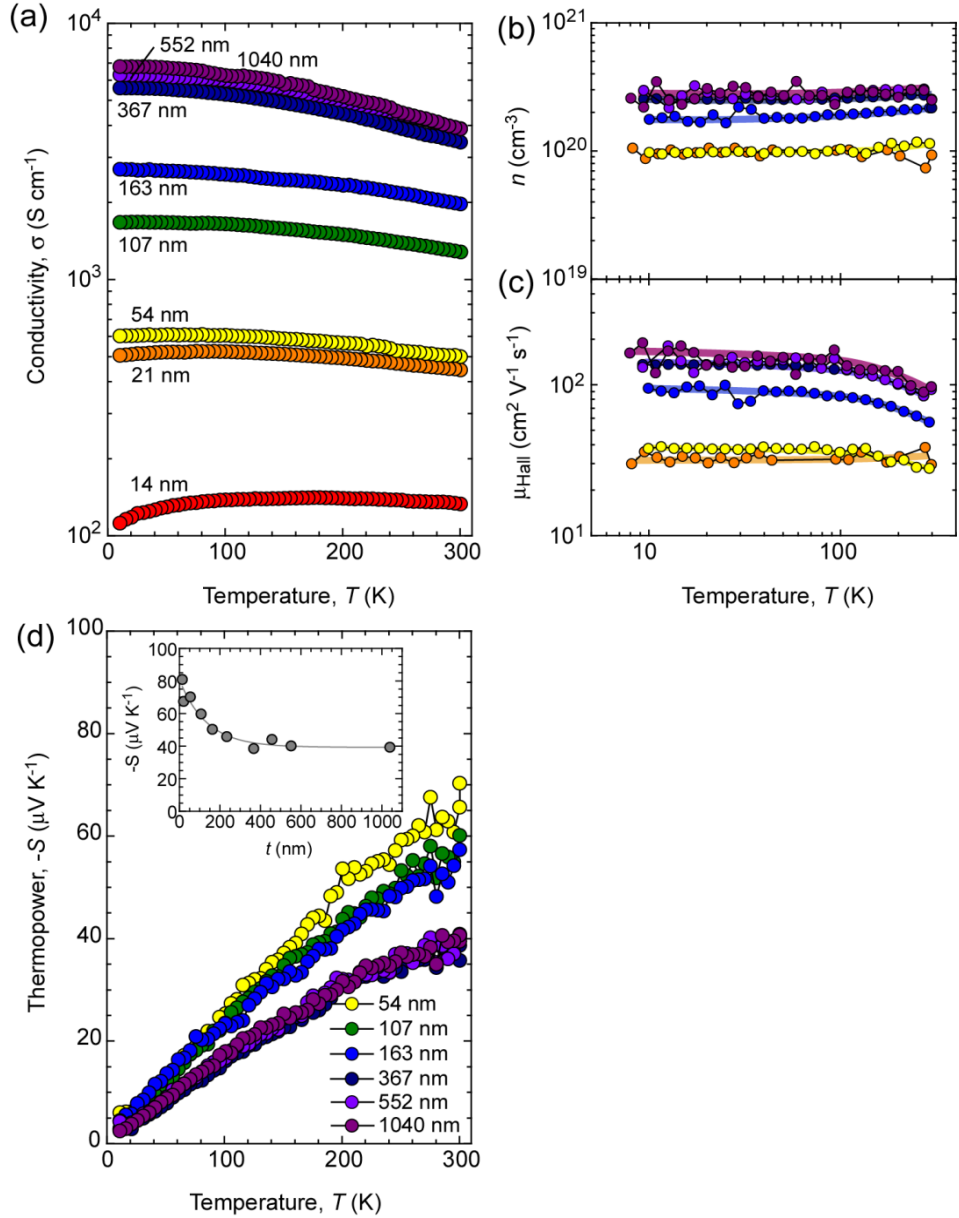

**Supplemental FIG. S4 | Temperature dependence of electron transport properties of the LBSO epitaxial films grown on (001) SrTiO$_3$ substrate.** [(a) conductivity ($\sigma$) (log scale), (b) carrier concentration ($n$) (log scale), (c) Hall mobility ($\mu_{Hall}$) (log scale), and thermopower ($S$)].

**Supplementary Table I** | Thickness dependent electron transport properties of the LBSO epitaxial films grown on (001) SrTiO$_3$ substrate at RT. ($\sigma$: electrical conductivity, $n$: carrier concentration, $\mu_{Hall}$: Hall mobility, $S$: thermopower)

| Thickness (nm) | $\sigma$ (S cm$^{-1}$) | $n$ (cm$^{-3}$) | $\mu_{Hall}$ (cm$^2$ V$^{-1}$ s$^{-1}$) | $S$ ($\mu$V K$^{-1}$) |
|---|---|---|---|---|
| 14 | 118 | 4.90×10$^{19}$ | 15 | −80.9 |
| 21 | 443 | 9.30×10$^{19}$ | 29.7 | −67.4 |
| 54 | 488 | 8.79×10$^{19}$ | 34.7 | −70.2 |
| 107 | 1211 | 1.32×10$^{20}$ | 57.2 | −59.7 |
| 163 | 1866 | 1.66×10$^{20}$ | 70.2 | −50.3 |
| 234 | 2393 | 1.83×10$^{20}$ | 81.8 | −45.8 |
| 367 | 3373 | 2.18×10$^{20}$ | 96.8 | −38.5 |
| 457 | 2632 | 1.72×10$^{20}$ | 95.8 | −44.1 |
| 552 | 3802 | 2.54×10$^{20}$ | 93.6 | −40.1 |
| 1040 | 3918 | 2.51×10$^{20}$ | 97.7 | −39.3 |

**Supplementary Table II** | Thickness dependent electron transport properties of the LBSO epitaxial films grown on (001) MgO substrate at RT. ($\sigma$: electrical conductivity, $n$: carrier concentration, $\mu_{Hall}$: Hall mobility, $S$: thermopower)

| Thickness (nm) | $\sigma$ (S cm$^{-1}$) | $n$ (cm$^{-3}$) | $\mu_{Hall}$ (cm$^2$ V$^{-1}$ s$^{-1}$) | $S$ ($\mu$V K$^{-1}$) |
|---|---|---|---|---|
| 18 | 360 | 8.86×10$^{19}$ | 25.4 | −51.4 |
| 44 | 1327 | 1.64×10$^{20}$ | 50.6 | −38.3 |
| 83 | 2065 | 1.87×10$^{20}$ | 68.8 | −35.6 |
| 95 | 2698 | 2.04×10$^{20}$ | 82.46 | −32.6 |
| 151 | 2470 | 1.98×10$^{20}$ | 77.7 | −34.9 |
| 220 | 3440 | 2.41×10$^{20}$ | 89.2 | −32.3 |
| 302 | 3328 | 2.31×10$^{20}$ | 89.9 | −34.5 |
| 350 | 3720 | 2.52×10$^{20}$ | 91.99 | −33.4 |
| 400 | 2670 | 1.86×10$^{20}$ | 89.8 | −31.8 |
| 450 | 3720 | 2.34×10$^{20}$ | 99.2 | −33.9 |
| 800 | 3187 | 2.10×10$^{20}$ | 94.7 | −34.5 |
| 1000 | 3088 | 2.18×10$^{20}$ | 88.52 | −34.8 |